**Google or ChatGPT: Who is the Better Helper for University Students**

Mengmeng Zhang [a], Xiantong Yang [b,*]

[a] School of Education, Minzu University of China, Beijing 100081, China

[b] Faculty of Psychology, Beijing Normal University, Beijing 100875, P.R. China

**Contact Information for all authors:**

Mengmeng Zhang: zhangmm817@foxmail.com

Xiantong Yang: xtyang93@foxmail.com (Corresponding Author)

# Abstract

Using information technology tools for academic help-seeking among college students has become a popular trend. In the evolutionary process between Generation Artificial Intelligence (GenAI) and traditional search engines, when students face academic challenges, do they tend to prefer Google, or are they more inclined to utilize ChatGPT? And what are the key factors influencing learners' preference to use ChatGPT for academic help-seeking? These relevant questions merit attention. The study employed a mixed-method research design to investigate Taiwanese university students' online academic help-seeking preferences. The results indicated that students tend to prefer using ChatGPT to seek academic assistance, reflecting the potential popularity of GenAI in the educational field. Additionally, in comparing seven machine learning algorithms, the Random Forest and LightGBM algorithms exhibited superior performance. These two algorithms were employed to evaluate the predictive capability of 18 potential factors. It was found that GenAI fluency, GenAI distortions, and age were the core factors influencing how university students seek academic help. Overall, this study underscores those educators should prioritize the cultivation of students' critical thinking skills, while technical personnel should enhance the fluency and reliability of ChatGPT and Google searches, and explore the integration of chat and search functions to achieve optimal balance.

Keywords: Academic help-seeking, Machine learning, ChatGPT, Google, University students

**1 Introduction**

The use of information technology tools for seeking help has become a common practice among college students, drawing widespread attention from scholars (Amador & Amador, 2014; Cheng et al., 2013; Fan & Lin, 2023). Traditional interpersonal help-seeking, such as reaching out to teachers and classmates for academic assistance, may have drawbacks. For example, research suggests that students may feel stressed or perceive a threat to their self-esteem when publicly sharing certain needs (Almed et al., 2017; Peeters et al., 2020). In contrast, seeking help from a machine provides flexible, convenient, real-time, and personalized support while safeguarding personal privacy, ultimately alleviating stress related to interpersonal relationships (Adams et al., 2023; Fan & Lin, 2023; Giblin et al., 2021).

With the development of Generation Artificial Intelligence (GenAI), the number of students using GenAI is on the rise, which is changing the way college students explore new things (Kasneci et al., 2023). Unlike traditional web search engines (e.g., Google), GenAI such as ChatGPT is built on deep learning and can learn and generate human-like responses. It can not only respond but also generate related content based on subsequent questions and prompts derived from the initial response (Sun et al., 2022). Tailored to educational settings, learners believe that ChatGPT can be used as a learning tool to provide personalized and immediate learning support and feedback (Adams et al., 2023; Ding et al., 2023; Kuhail et al., 2023). However, the accuracy of the information provided by ChatGPT is concerning (Kidd & Birhane, 2023). Compared to traditional search engines, GenAI may struggle with producing fantastical or nonsensical content, potentially compromising the accuracy of the information provided (Xu et al., 2023), and potentially impacting students' preference to use them for help-seeking.

Based on the above studies, when students face academic challenges in the evolutionary process between emerging technologies and traditional methods, do they prefer traditional search engines (e.g., Google), or are they more inclined to use GenAI tools like ChatGPT? (**research question one**). Additionally, according to the affordance-actualization theory (Gibson, 1977) and prior empirical research (Adams et al., 2023; Cheng & Tsai, 2011; Cheng et al., 2023), students' preferences for online academic help-seeking are influenced by both objective and subjective factors. Whether these factors also influence students' preferences for seeking help using ChatGPT or Google remains lacking in relevant empirical research. Thus, **research question two** is proposed: What are the core factors influencing learners' preference to use ChatGPT for academic help-seeking? Delving deeper into this topic is crucial for promoting the development of educational technology, enhancing the quality of academic support for students, and meeting their personalized needs.

## 1.1. Literature review

### 1.1.1 Theoretical framework

Online academic help-seeking involves spontaneously requesting assistance from others through the particular vehicle of the Internet. It specifically pertains to seeking assistance through online tools (e.g., traditional search engines like Google or GenAI tools like ChatGPT), aimed at solving academic problems (Fan & Lin, 2023). Relevant research suggests that students' academic-seeking behavior is a complex phenomenon influenced by multiple factors (Adams et al., 2023; Giblin et al., 2021). Affordance-actualization theory (Gibson, 1977) emphasizes the interaction between the individual and the environment, and how the individual actively perceives and uses opportunities in the environment to achieve their goals. Affordance is an action potential that an object or a situation offers to an agent. From an interactive perspective, affordance typically refers to the cues or properties of a technology that guide users on how to interact with technologies, serving as clues for operation (Shin, 2021). Actualization is an active process whereby individuals perceive and utilize affordances within objects or their context. This process involves individuals recognizing and selecting interactive opportunities provided by the surroundings through cognitive and emotional processes to achieve goals (Shin, 2022). The affordance-actualization theory posits that the objective characteristics of technology (Affordances Existence) and users' subjective perceptions (Affordances Perception) jointly influence actions regarding technology adoption (Affordances Actualization).

In some studies, researchers explore the interaction between technological attributes and user psychological characteristics, as well as their effects in similar usage environments (Ali et al., 2023; Jónasdóttir & Müller, 2020; Shin, 2022). For example, Ali et al. (2023) explored how personalized travel recommendations from ChatGPT influence travelers' behavioral intentions. Consideration of the users' behavior preference between Google and ChatGPT is not only influenced by the objective technical features but also by their subjective perceptions and understanding of the technology. The affordance-actualization theory was chosen as a theoretical framework for this study because it provides an effective perspective for connecting the objective features of technology and the subjective characteristics of users. This connection allows for an exploration of the factors that influence university students' preferences for seeking academic help.

1.1.2 Objective factors

The first group of factors focuses on objective factors. **Fluency** is an important factor in behavioral choice and judgment (Undorf et al., 2017). Higher fluency means easier and faster processing of information, which generally leads to more positive preferences (Shen et al., 2018). When a tool or platform provides smooth communication and problem-solving skills, users are more likely to credibility it (Kasneci et al., 2023; Lankes, 2008). One reason ChatGPT is highly favored is its ability to respond to conversations in a coherent, natural manner, making users feel understood and their problems addressed (Kasneci et al., 2023). Therefore, students with perceived higher ChatGPT fluency may be more inclined to choose ChatGPT for academic help-seeking.

**Accuracy** as the second objective factor may predict university students' choice regarding academic help-seeking. Prior study indicates that individuals who perceive ChatGPT as more accurate are more likely to use it for seeking help (Kim et al., 2023). Conversely, if students doubt ChatGPT accuracy, they may opt for other, more reliable resources, such as using search engines like Google. **Anthropomorphism** as the third objective factor provides users with the sensation of interacting with a "virtual person". Related research suggests that anthropomorphism enhances students' perceived social connection with AI and their willingness to adopt AI technologies (Ding et al., 2023).

1.1.3 Subjective factors

The second group of factors focuses on individual subjective factors. Individuals with ChatGPT **distortion** may overestimate their GenAI performance (Urban et al., 2024). Focusing on the context of online academic help-seeking, individuals with higher GenAI distortion are more likely to trust the aggregated information, structured responses, and intuitive explanations provided by ChatGPT (Church, 2024; Shoufan, 2023). Conversely, Google provides divergent information, requiring users to filter and extract information (Wu et al., 2024), which poses a greater challenge to individuals' cognition. **Cognitive reflection** as the second subjective factor refers to the ability to interrupt intuitive thinking and engage in a deliberate thought process at the appropriate moment (Frederick, 2005). Thus, individuals with higher cognitive reflection may use traditional search methods (e.g., Google) for thoughtful analysis and problem-solving rather than simply accepting AI-generated answers. Third, individuals with an **analytical cognitive style** tend to avoid errors in rational thinking and prefer replacing intuitive responses. They spend time considering alternatives and engage in extensive mental simulations to tackle new challenges (Viator et al., 2019). When facing academic challenges, they are less likely to use GenAI tools directly for answers. Instead, they prefer seeking detailed explanations and engaging in logical reasoning. Fourth, individuals who maintain a **skeptical** attitude are more likely to assess the accuracy of the answers provided by GenAI (Ahadzadeh et al., 2023; Buchanan & Hickman, 2023), allowing them to use the information generated by large language models more wisely without blindly accepting it (Rusandi et al., 2023). Fifth, individuals with **inert thinking** often hold a negative attitude towards deep thinking and proactive learning (Authors, under review). When they need academic assistance, they are more likely to choose tools that provide convenience and immediate responses (e.g., ChatGPT).

As the sixth subjective factor, Users who have a **positive affect towards ChatGPT** (e.g., perceived as "warmth") are more likely to increase their emotional trust in AI and are more willing to establish and maintain effective connections with the AI system (Ding et al., 2023; Gonzalez-Jimenez, 2018). Regarding **expectation beliefs** as the seventh subjective factor, individuals who hold positive expectation beliefs about behavior are more likely to engage in that behavior (Panitz et al., 2021). In academic help-seeking, positive expectations about ChatGPT may drive users to greater usage, and acceptance of its answers.

Self-leadership as the eighth subjective factor is defined as a process of self-influence through which an individual can achieve self-motivation and self-direction, thereby enhancing their performance (Hauschildt & Konradt, 2012). Self-leadership stresses individuals' proactive control over their actions, encompassing goal setting and monitoring one's performance (Hauschildt & Konradt, 2012). In the GenAI context, those adept at **self-leadership** may prefer using search engines like Google to tackle challenges. This preference stems from search engines offering a vast array of information resources, enabling them to independently search for and sift through the information they require, thereby enhancing their ability to manage their learning process effectively. Individuals with higher **learning avoidance motivation** as the ninth subjective factor may tend to opt for simple and convenient methods to complete tasks, rather than investing effort in independent learning (Ye et al., 2023). When facing academic challenges, these individuals are more likely to prefer using efficient assistance tools such as ChatGPT to easily obtain answers without the need for extensive thought or effort. **Accuracy motivation** as the tenth subjective factor refers to the individuals' drive to obtain accurate and reliable information (Rathje et al., 2023). In the choice of academic help-seeking, college students tend to choose tools they believe can provide accurate answers. If tools like ChatGPT or Google are perceived as accurate and reliable sources, students are more likely to use them to address their academic inquiries.

1.1.4 Demographic factors

Considering that individual background may also predict behavioral preference, we included individual information in the last group of factors. Studies focusing on **sex** have shown that girls are more active and more adaptive in academic help-seeking behaviors than boys (Cheng et al., 2023). Regarding **age and grade**, younger people's preference for chatbots over older ones aligns with the general trend of adopting new technology (Thormundsson, 2023). Compared to students lacking **experience** in ChatGPT, users with experience using ChatGPT and a high frequency of use are more likely to seek ChatGPT's assistance in completing academic tasks (Adams et al., 2023).

1.1.5 Machine learning algorithms

Previous academic help-seeking studies primarily used classical statistical methods such as logistic regression (Adams et al., 2023; Cheng & Tsai, 2011). However, simple regression techniques may potentially oversimplify predictor outcome associations and reduce prediction accuracy (Yarkoni & Westfall, 2017). Moreover, classical statistical analysis fails to rank factors by predictor variable importance, unlike machine learning algorithms. These algorithms excel in learning complex, nonlinear relationships and evaluating factors' impact through methods like feature importance analysis (Sen et al., 2021). Furthermore, machine learning algorithms can handle massive datasets, unlike classical models with strict data requirements (Breiman, 2001). They can identify potential nonlinear associations and make more accurate predictions regarding university students' academic help-seeking. In logistic regression models with some predictors, multicollinearity issues may arise, making it challenging to compare predictors' relative importance based on standardized regression coefficients (Lavrijsen et al., 2022). However, machine learning algorithms typically offer higher prediction accuracy by uncovering nonlinear associations in the learning data. Therefore, considering the limitations of traditional statistical methods, this study, based on GenAI, employs the optimal machine learning model to examine the predictive role of these factors in Taiwanese university students' academic help-seeking choices.

## 1.2. The present study

To the best of our knowledge, no research has yet compared and evaluated the best-performing machine learning models for detecting the key factors influencing university students' choices regarding academic help-seeking. Examining college students' views on GenAI and employing machine learning models to identify the key factors influencing their online academic help-seeking is crucial to understanding user psychological behavior preferences. This investigation facilitates precise optimization of language model algorithms, enhancing the model's reliability (Kasneci et al., 2023), and tailoring it more closely to users' preferences and needs.

Considering the use of mixed methods allows for a comprehensive exploration of the research objectives by integrating quantitative and qualitative data collection and analysis techniques (Creswell, 2021; Tashakkori et al., 2020). This study employed a sequential explanatory mixed-method research design to investigate university students' preference for online academic help-seeking. This research has two main objectives. Firstly, this study aims to investigate university students' preferences between Google and ChatGPT when seeking academic assistance. Given the acceptance of new technologies among university students (Adams et al., 2023), we hypothesize that they prefer using ChatGPT for academic help-seeking. Secondly, based on affordance-actualization theory (Gibson, 1977) and previous studies, we comprehensively examine the predictive effects of different factors on preferences for academic help-seeking, aiming to identify the strongest predictive factor. Considering the data-driven approach, we cannot provide explicit hypotheses.

## 2. Methods

### 2.1. Participants

Given the unrestricted access to ChatGPT and Google in Taiwan, this study aims to explore Taiwanese university students' preferences in using two online tools for academic assistance. The quantitative data consisted of 916 college students who completed data collection. There were 598 (65.3%) male and 318 (34.7%) female. Among them, there were 553 undergraduate students, 320 master's students, and 43 doctoral students. The age ranges from 18 to 34 years old, with an average age of 22.65 ($SD$ = 3.19). There were no systematic differences between participants who attended all measurement occasions and those who didn't. All procedures were approved by the Institutional Review Board of the first author.

In qualitative data, semi-structured interviews have been developed for this study. Recruit research subjects through a public platform (https://www.credamo.com), following the principle of voluntariness. Based on the time availability of participants, we scheduled meeting times. The online interview duration (Zoom/Tencent Meeting) ranged between 15 and 30 minutes, and it was organized using a non-probability, purposive approach. At the beginning of the interview, the researcher explained the purpose and content of the interview to the 8 participants. Subsequently, we introduced the topic to the participants and posed several questions to obtain the participants' opinions. The conversations were documented and transcribed. Interviews continued until reaching saturation standards (Malterud et al., 2021). Appendix A1 shows the demographic information of the participants.

## 2.2. Measures

Influence factors in this study include university students' individual information (e.g., sex, age, grade, use experience), subjective factors (ChatGPT distortions, cognitive reflection, analytical cognitive style, skepticism, inert thinking, positive affect towards ChatGPT, expectation beliefs, self-leadership, learning avoidance motivation, accuracy motivation), and objective factors (ChatGPT fluency, ChatGPT accuracy, anthropomorphism). Regarding the dependent variable, the study employed one question to evaluate the preferred type of academic help-seeking among college students. The question asked, "When I encounter difficulties in my studies, I prioritize seeking help from A. Google or B. ChatGPT." Given that university students have primarily used search engines for assistance in the past, we used Google as a reference point to examine the preferences of university students when seeking academic help using Google or ChatGPT. We considered Google as the baseline for seeking help, encoded as 0, and ChatGPT encoded as 1. Supplementary materials (Appendix A3) provide detailed descriptions of each variable and sample item.

After conducting quantitative research, a qualitative survey method was employed. This involved using some interview questions such as, "Why do you prefer using ChatGPT or Google when facing academic difficulties?" (see Appendix A2) to gather participants' reasons for preferences regarding Google or ChatGPT when seeking academic assistance.

## 2.3. Analytical approach

Regarding the quantitative phase, to enhance the original dataset's resolution, we utilized the SMOTE algorithm (Chawla et al., 2002) for oversampling, increasing the number of minority class members in the training set. Unlike down-sampling, the oversampling technique retains all members from both minority and majority classes in the original training set. In this study, a larger over-sampled dataset comprising 948 participants was generated, with 663 respondents assigned to the training group and 285 respondents to the testing group.

Subsequently, seven machine learning techniques, namely Logistic Regression, Naive Bayes, Decision Tree, Random Forest, K-Nearest Neighbors (KNN), Light Gradient Boosting Machine (LightGBM), and Artificial Neural Network (ANN) are employed to assess and compare model performance. The core characteristics of each algorithm have been provided in the supplementary materials (Appendix B). Each model comprises 18 variables representing factors that influence university students' preference for academic help-seeking behaviors. A series of metrics (Accuracy, Precision, Specificity, Recall, and F1-score) were used to validate the algorithm performances. After identifying the optional algorithms based on different model evaluation metrics, the feature importance factors that predicted university students' preference for academic help-seeking were shown and discussed.

Regarding the specific meanings of metrics, the Accuracy metric typically represents the degree of consistency between predicted values and actual values (Hussain et al., 2019). In the context of university students' academic help-seeking, Accuracy helps evaluate the predictive capability of the model, i.e., whether the model accurately predicts the types of academic help-seeking preference among university students. Precision identifies the probability of a positive test result. High precision values indicate that the probability of the test set being accurately classified will be high. Recall evaluates the number of true positives of the actual class predicted by the models (Sweeney et al., 2016). Higher Recall scores indicate better classifier performances. The F1 score combines precision and recall into a single value, providing a balanced measure of a model's overall performance. Besides, ROC (Receiver Operating Characteristics) and AUC (Area

Under Curve) were also used to assess the performance of binary classification models. They provided a comprehensive view of how well a model can distinguish between positive and negative cases (Yan et al., 2023). In this study, the higher AUC score indicates the algorithm's effectiveness in distinguishing between the "Google help-seeking" and the "ChatGPT help-seeking".

After selecting the best-performing algorithm, we used the LightGBM algorithm to 'split' feature importance. Identifying the relative importance of features helps to determine the key predictors influencing academic help-seeking preference among university students. To further assess which feature plays the most crucial role in determining academic help-seeking preference among university students, we conducted a 10-fold cross-validation along with the application of SHAP methods (Shapley Additive exPlanations). These methods were employed to enhance the interpretability of machine learning models (Lundberg and Lee, 2017). Specifically, SHAP values quantify the impact of each feature on the model prediction. Positive SHAP values indicate that a feature contributed to pushing the prediction higher, while negative values suggest the opposite.

During qualitative analysis, we initially reviewed participants' online questionnaire responses to understand their ChatGPT and Google usage patterns, providing insights into their backgrounds. Thematic analysis, following Braun and Clarke (2022), was then conducted, where two researchers identified patterns, themes, and key concepts in the interview data. These encapsulated participants' perspectives, preferences, and influential factors. Finally, the qualitative findings were utilized to scrutinize, explore, and interpret the phenomena uncovered in the quantitative research, along with the underlying reasons.

## 3. Results

### 3.1. Students' preference to use ChatGPT or Google for academic help-seeking

In the entire sample, 442 students (48.3%) preferred using Google for academic help-seeking, while 474 students (51.7%) preferred using ChatGPT for the same purpose. Overall, the usage rates of both types of tools were relatively high. Comparatively, students tended to prefer using ChatGPT to seek academic help, reflecting the potential application and popularity of GenAI in the educational field. From interviews with eight students, we discovered unanimous agreement regarding the effectiveness of ChatGPT in aiding academic learning. All eight students believed that utilizing ChatGPT can enhance problem-solving efficiency. However, some students also voiced concerns, stating that ChatGPT occasionally offers misleading answers (Participants 5, 6, 7, 8). As a result, they often depended on Google to search for more objective information.

"*When I write papers, I prioritize rigor. In my daily writing, if I come across issues like awkward sentence structure, I usually turn to ChatGPT for assistance. However, when I need to verify factual viewpoints, I prefer using Google to search for literature*." (Participant 8).

Table 1 shows the descriptive statistics. We found that most variables were significantly predicting students' online help-seeking preferences. Meanwhile, the results of $t$-tests indicated significant differences in academic help-seeking preference between Google and ChatGPT across most indicators. This suggested that users exhibit different responses when seeking different sources of assistance.

Table 1. Descriptive statistics of the variables

| variable name | Total sample (*N*=916) | | | Google help-seeking (*N*=442) | | ChatGPT help-seeking (*N*=474) | | The *t*-test between different help-seeking |
|---|---|---|---|---|---|---|---|---|
| | *M* | *SD* | *r* | *M* | *SD* | *M* | *SD* | *t*-value |
| sex | 1.35 | 0.48 | -0.05 | 1.37 | 0.48 | 1.32 | 0.47 | 1.46 |
| age | 22.65 | 3.19 | .11** | 22.29 | 3.37 | 22.99 | 2.96 | -3.34 |
| grade | 1.44 | 0.58 | 0.04 | 1.42 | 0.58 | 1.46 | 0.59 | -1.12 |
| use experience | 2.91 | 0.82 | .10** | 2.83 | 0.73 | 2.99 | 0.88 | -3.06*** |
| self-leadership | 3.93 | 0.52 | .12** | 3.86 | 0.57 | 3.99 | 0.47 | -3.70*** |
| GenAI distortion | 3.55 | 0.73 | .15** | 3.44 | 0.74 | 3.66 | 0.71 | -4.52*** |
| analytic cognitive style | 3.26 | 1.06 | -0.01 | 3.26 | 1.08 | 3.25 | 1.05 | 0.13 |
| accuracy motivation | 3.60 | 1.09 | 0.02 | 3.58 | 1.06 | 3.63 | 1.11 | -0.66 |
| skepticism | 3.25 | 0.99 | -.11** | 3.36 | 0.92 | 3.15 | 1.04 | 3.34*** |
| positive affect | 3.91 | 0.54 | .07* | 3.88 | 0.55 | 3.95 | 0.53 | -2.03* |
| inert thinking | 3.38 | 0.77 | -0.06 | 3.43 | 0.77 | 3.34 | 6.13 | 1.89 |
| expectancy beliefs | 3.89 | 0.50 | .13** | 3.82 | 0.53 | 3.95 | 0.46 | -4.01*** |
| avoidance motivation | 3.07 | 1.00 | -.12** | 3.19 | 0.97 | 2.96 | 1.01 | 3.52*** |
| anthropomorphism | 3.24 | 0.54 | -0.04 | 3.26 | 0.51 | 3.21 | 0.57 | 1.48 |
| GenAI accuracy | 3.87 | 0.55 | .10** | 3.81 | 0.57 | 3.92 | 0.51 | -2.90*** |
| GenAI fluency | 3.43 | 0.46 | .14** | 3.36 | 0.49 | 3.49 | 0.42 | -4.19*** |
| cognitive reflection | 1.34 | 0.82 | .11** | 1.44 | 0.76 | 1.25 | 0.86 | 3.47*** |

Note: *p < 0.05; **p < 0.01; ***p < 0.001.

## 3.2. The key predictors of academic help-seeking preference among university students

All 18 variables in the dataset were used as input features in a machine-learning model. After testing and comparing seven machine learning algorithms, we found that Random Forest and LightGBM exhibit notable algorithmic advantages in evaluating the importance of each factor in predicting academic help-seeking preference. To enhance the accuracy and performance of the model, we integrated feature importance from LightGBM and Random Forest algorithms into a combined model for accurately predicting university students' preferences between Google and ChatGPT for academic help-seeking. Regarding the process of comparing and identifying machine learning models, we've outlined in supplementary materials (Appendix C). This section predominantly showed the prediction outcomes of machine learning.

Fig. 1 displays the top significant features of academic help-seeking preferences among university students, respectively. The higher the feature importance valued, the more important the feature was within this machine learning model. Combining the Random Forest and the LightGBM algorithms, the average values of feature importance were computed in Fig. 2 (3). Specifically, GenAI distortions (CDS) were found to be the most relevant factor influencing university students' preferences for obtaining information. Additionally, GenAI fluency (CF) and age emerged as other top variables in the analysis.

Fig. 2 presents the summary plots of SHAP graphs that combine feature importance with feature effects. The chart illustrated the directional effects of different factors at varying levels. As feature values increased, the color of the dot tended to become redder, while as feature values decreased, the color of the dot leaned towards bluer. The larger the SHAP value was, the greater its influence on the model's output was. The results indicated that higher levels of ChatGPT fluency (CF) and ChatGPT distortions (CDS) are associated with an increased tendency among university students to seek help from ChatGPT. University students of higher age tended to show a greater inclination toward using traditional information search tools (i.e., Google).

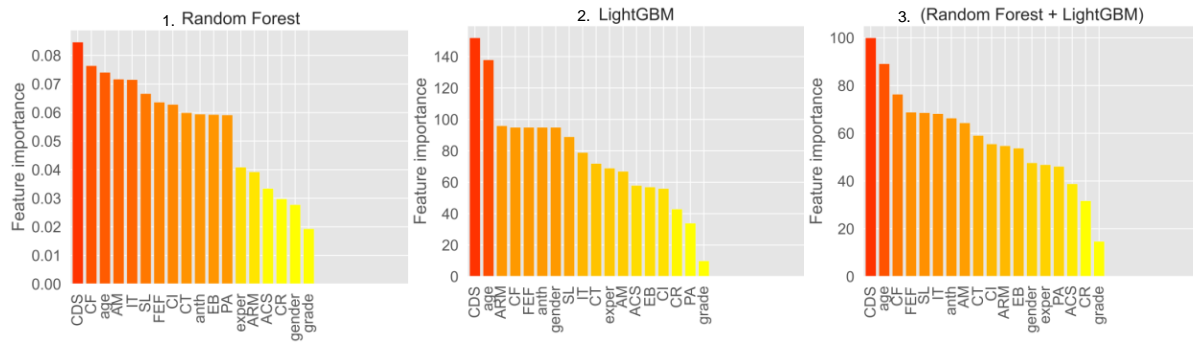

**Fig. 1.** Top significant features for academic help-seeking preferences among university students

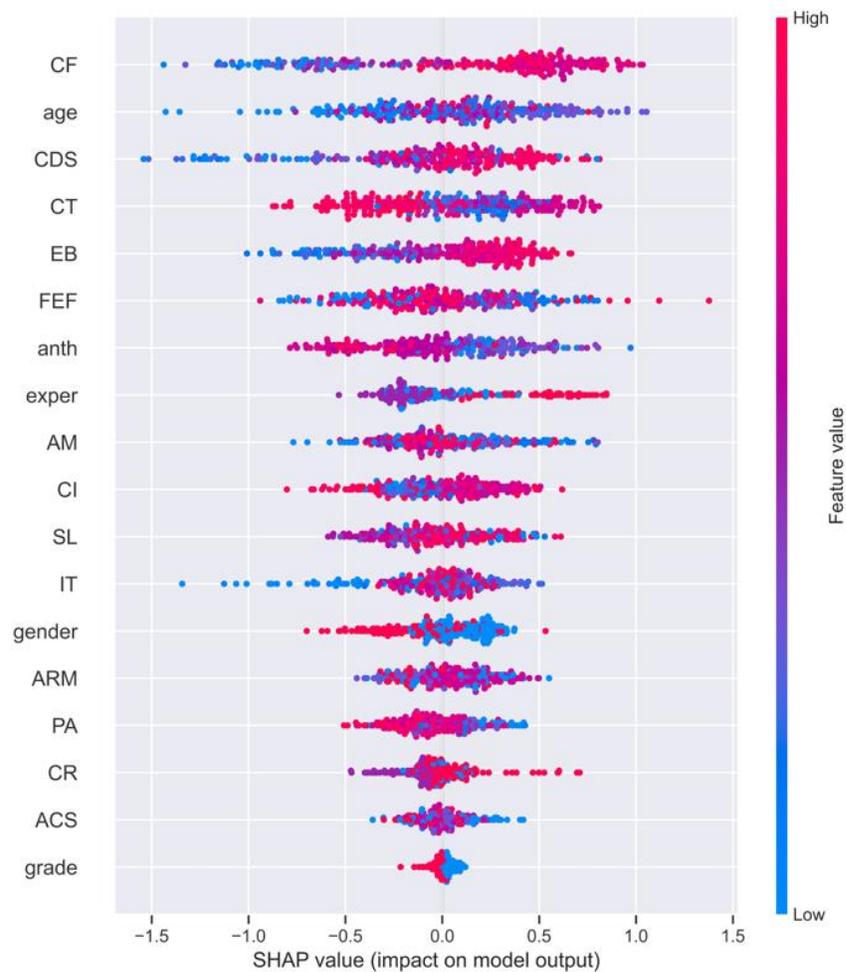

**Fig. 2.** SHAP values for academic help-seeking preference among university students

Findings from the semi-structured interview supported the quantitative results. Among the eight students interviewed, they all expressed their willingness to use ChatGPT for academic assistance, especially when dealing with complex and obscure topics. Their feedback highlighted the friendliness and intelligence of ChatGPT. In the students' responses, a recurring theme was the convenience and fluidity of their conversations with ChatGPT. They generally believed that ChatGPT could provide natural and fluent responses to their inquiries. This contrasted with the need for manual filtering and discernment when using traditional search engines to obtain answers. One student expressed,

"*When I use ChatGPT to search for information, it always promptly responds to my queries. In contrast, when using Google for information searches, I can some results that require me to spend time actively filtering to find the information I need.*" (Participant 1).

"*When writing reports or essays, I sometimes encounter situations where the language expression is unclear, and I don't know how to organize the language better. I input the content that needs modification into ChatGPT, and it can fluently provide me with new expressions while maintaining the same content meaning.*" (Participant 4).

Through machine learning predictions, we also found that GenAI distortions predict students' preferences for ChatGPT when seeking academic help. Previous research has shown that the GenAI model can distort individuals' beliefs since it can't differentiate between facts and fiction (Kidd & Birhane, 2023). Users who overestimate GenAI's response tended to trust GenAI more and thus seek its assistance more often. Besides, in interviews, one student mentioned,

"*Although ChatGPT may occasionally make mistakes, its vast knowledge base surpasses mine, and I consider its answers authoritative. It's like having a reliable assistant continually helping me.*" (Participant 3).

However, students, when dealing with complex questions, were inclined not to readily trust the answers provided by ChatGPT and instead adopt a multi-channel verification approach, such as using a combination of Google and ChatGPT for cross-validation. One student stated,

"*When I use ChatGPT to inquire about certain uncertain questions, although its answers sound very smooth, my rationality tells me that I must seek more evidence to support these answers. Therefore, I will switch to using Google for further verification*." (Participant 7).

Based on machine learning predictions, older students tended to prefer using Google search compared to younger students. This inclination may be related to the academic needs of students in different age groups. In qualitative interviews, two older doctoral students expressed a preference for Google, citing its utility in accessing literature and verifying academic content (Participants 6 and 8). However, it's important to clarify that they do not exclude the use of ChatGPT. When facing explanatory questions, they often found that ChatGPT provide answers that are easier to understand and more accessible. Therefore, age differences may predict students' preferences for academic assistance tools, reflecting varying academic requirements across age groups.

In summary, the findings of qualitative research further explain the results of quantitative research. The study emphasized that ChatGPT fluency, cognitive distortion, and age were core factors influencing preferences for seeking academic assistance. Nonetheless, students did not necessarily prefer a specific way of academic assistance. In different problem contexts, combining the use of ChatGPT and Google was more likely to objectively and accurately address issues.

## 4. Discussion

This study first investigates university students' preferences between Google and ChatGPT when seeking academic assistance. We found that the usage rates of ChatGPT and Google are relatively high. In comparison, students generally prefer using ChatGPT for academic assistance driven by the perceived enhancement of problem-solving efficiency it offers. Notably, while ChatGPT offers convenient question-answering and language processing features, the Google search engine provides a wider array of information sources. It helps students compare and evaluate various viewpoints and resources, leading to a more comprehensive understanding of specific topics or issues. Therefore, when seeking academic assistance through either ChatGPT or Google, students need to develop critical thinking skills and be mindful of assessing the accuracy and reliability of information to acquire comprehensive and precise knowledge.

After identifying the most powerful algorithms (Random Forest and LightGBM), this study utilized the affordance-actualization theory to comprehensively explore the predictive effects of different factors on university students' academic help-seeking preferences. In general, our study indicated that the objective characteristics of technology (e.g., ChatGPT fluency) and users' subjective perceptions (e.g., ChatGPT distortions) jointly influence actions regarding technology adoption (e.g., academic help-seeking preferences). This insight not only confirms the validity of the affordance-actualization theory but also extends its application to the field of GenAI, thereby broadening the theory's scope of applicability.

Specifically, we found that ChatGPT fluency has been demonstrated as one of the factors associated with university students' preferences for academic help-seeking. This finding aligned with prior studies, showing that ChatGPT responds to conversations in a coherent, natural way, making users feel understood and their problems solved (Kasneci et al., 2023; Lankes, 2008), thus users are more likely to use it.

Moreover, university students with higher levels of ChatGPT distortions are more likely to use ChatGPT for academic help-seeking. Excessive trust is a specific manifestation of ChatGPT distortions (Authors, under review). Users who trust GenAI tend to overestimate ChatGPT and seek its assistance more often (Urban et al., 2024). Church's (2024) research also confirmed that some students excessively trust responses from ChatGPT during assignments, even if they're fabricated. This is because subjectively believing in ChatGPT is easier than rational skepticism while fact-checking and internet searches are seen as cumbersome. The above conclusion implies the significant role of GenAI distortion in individual behavioral preferences. This has also sparked concerns about how to effectively guide students to use traditional and GenAI assistants more rationally and cautiously.

Finally, older university students tend to prefer using Google, while younger students may be more inclined to use ChatGPT. Related statistics found that ChatGPT is most widely used among the global population aged 25 to 34. The second-largest user group comprises individuals under 24, with those under 34 constituting over 60% of ChatGPT users. The preference for chatbots among younger individuals over older ones aligns with the general trend of new technology adoption (Thormundsson, 2023). Through interview materials, this study explains from the perspective of students' academic needs. The survey sample covers undergraduate, master's, and doctoral students. Older students typically have higher academic qualifications and more demanding academic tasks, so they often prefer using search engines like Google to obtain reliable materials and evidence when seeking academic assistance. In comparison, although ChatGPT may be reliable in certain explanatory answers, its answers may lack sufficient evidence (Xu et al., 2023) and be unreliable in academic information retrieval.

## 5. Theoretical and practical implications

Theoretically, our study extends the application of affordance-actualization theory into the realm of GenAI. While previous studies primarily focused on its application in general technologies (e.g., information systems; Leonardi, 2011), the advent of GenAI characterized by fluency response (Kidd & Birhane, 2023) and high anthropomorphism (Alabed et al., 2023) raises questions about whether this theory can explain human-computer interaction preferences within the GenAI context. Our study provides empirical insights into this question.

Practically, this study may help technology developers and educators use relatively fewer indicators (such as ChatGPT fluency, ChatGPT distortions, and age) to understand whether university students prefer ChatGPT or Google when seeking academic assistance. For technology developers, this emphasizes that whether it's traditional search engines or GenAI tools, enhancing their fluency and content reliability is one of the ongoing improvement goals, enabling them to handle and manage complex tasks more intelligently (Shen et al., 2023).

Individuals with ChatGPT distortions tend to prefer using ChatGPT for academic assistance, yet they may struggle with objectively evaluating the accuracy of information provided by ChatGPT. Consequently, educators and universities should focus on nurturing students' critical thinking skills, empowering them to effectively discern the authenticity of feedback provided by artificial intelligence technologies (as cited in Adams et al., 2023), to confront the challenges brought about by GenAI. This research also contributes to students' readiness to confront the challenges posed by GenAI. For example, individuals with ChatGPT distortions prefer ChatGPT academic help-seeking, but they may struggle to assess the accuracy of the information provided by ChatGPT objectively. Therefore, teachers and universities need to foster students' critical thinking skills to enhance their ability to discern the authenticity of feedback provided by AI technologies (Adams et al., 2023).

It is important to acknowledge that, as revealed in our interviews, in an era where traditional and GenAI coexist, relying solely on one tool often poses certain limitations. While Google offers abundant information resources, users must actively sift through irrelevant content. ChatGPT stands out for its ability to provide immediate feedback, yet the accuracy of its responses may be questionable. To improve the robustness and user-friendliness of these tools, future research could explore integrating chat and search functionalities and finding the optimal balance between conversational and keyword-based retrieval methods (Xu et al., 2023).

## 6. Limitations and future research

Several limitations still need to be discussed. First, this study focused on Taiwanese university students, due to data limitations. Therefore, the research objects should be expanded to include university students from other regions. Second, based on the framework of affordance and actualization theory (Gibson, 1977), we have included some predictive factors in this study. Future studies should consider including more related variables (e.g., privacy and security concerns), that may have been overlooked in our analysis. Furthermore, this study mainly explored the direct predictive effects of exploratory variables on preferences for online academic help-seeking. There may be even more complex relationships among variables, such as mediation and moderation effects, which warrant further investigation in future research (Ding et al., 2023). Finally, this study primarily utilized cross-sectional data to examine predictive effects. Cross-sectional studies cannot validate specific causal relationships. Future research could verify the causal relationship between predictive factors and preferences for online academic help-seeking through experimental manipulation or longitudinal study.

## 7. Conclusion

This study employed a mixed-method research design to investigate university students' preference for online academic help-seeking. Firstly, we found that Taiwanese university students prefer using ChatGPT for academic help-seeking, but they also rely on Google for more objective information due to ChatGPT occasionally providing misleading answers. Secondly, we evaluated and compared seven machine learning algorithms in detecting key factors influencing academic help-seeking preference among university students. The study indicated that Random Forest and LightGBM algorithms have better performance in predicting university students' preferences for online academic help-seeking. Moreover, three factors that strongly predict university students' preference for academic help-seeking were identified: ChatGPT fluency, ChatGPT distortions, and age. Additionally, we employed qualitative analysis to delve deeper into the results uncovered by machine learning, aiming to uncover underlying reasons and provide additional insights. In summary, this study suggests that educators prioritize the cultivation of students' critical thinking skills, while technical personnel enhance the reliability of ChatGPT and Google searches, and explore the integration of chat and search functions to achieve optimal balance.

# References

Author et al. (2024)

Author et al. (2024)

**Supplementary materials**

**Appendix A1. Description of the interviewee**

| No. | Sex | Age | Major | Educational level |
|-----|--------|-----|------------------------|-------------------|
| 1 | Female | 22 | English Linguistics | undergraduate |
| 2 | Male | 21 | Preschooler Education | undergraduate |
| 3 | Female | 25 | Business Administration | Bachelor |
| 4 | Female | 24 | Educational Psychology | Bachelor |
| 5 | Female | 25 | Social work | Bachelor |
| 6 | Female | 28 | Higher Education | PhD |
| 7 | Male | 31 | Educational Psychology | PhD |
| 8 | Male | 30 | Geochemistry | PhD |

**Appendix A2. The interview questions**

1. Can you share your age, gender, and academic field?

2. How familiar are you with ChatGPT and Google?

3. Do you prefer using ChatGPT or Google when facing academic difficulties? Why do you prefer using ChatGPT or Google when facing academic difficulties?

4. Under what circumstances do you prefer using ChatGPT or Google? Why?

5. How effective do you find ChatGPT and Google in addressing academic challenges? Do they meet your expectations?

6. Which tool do you find more user-friendly for academic problem-solving, ChatGPT or Google? Are there any usage difficulties or limitations?

7. What factors do you consider when deciding to use ChatGPT or Google? For example, accuracy, fluency, anthropomorphism, etc.

8. In your opinion, what areas can ChatGPT and Google improve upon in solving academic problems?

Appendix A3. Detailed descriptions of each variable and sample item.

| Domain | Variables | Description/Sample item (Scale) |
|---|---|---|
| Objective factors | ChatGPT fluency | Based on prior research (Authors et al., under review), we adapted the ChatGPT fluency scale, which consists of 8 questions. The sample item is "ChatGPT answers questions smoothly." |
| | ChatGPT accuracy | We have developed an accuracy assessment scale for ChatGPT, comprising 8 questions. An example statement is "ChatGPT's responses are trustworthy." |
| | anthropomorphism | Adapted from the Anthropomorphism scale, based on the research by Bartneck et al. (2009). The sample item is "ChatGPT is conscious." |
| Subjective factors | ChatGPT distortions | Based on relevant research, we adapted the ChatGPT distortion scale (Authors, under review), which consists of 8 questions. e.g, "The answers provided by ChatGPT are authoritative." |
| | cognitive reflection | The cognitive reflection test (CRT) (Frederick, 2005; Pennycook et al., 2016) was used to test cognitive reflection. It is composed of three mathematical questions that each has a fast and intuitive, yet erroneous answer. For example, a 'lily pad' problem. |
| | analytical cognitive style | It was adapted from the study by Arechar et al. (2023). A sample item is "Compared to mentally demanding work, I prefer to do tasks that don't require much brainpower." |
| | inert thinking | Adapted from the inert thinking during ChatGPT, based on the research by Ye et al. (2024). A sample item is "I don't like delving into the causes of problems". |
| | skepticism | Adapted from the questionnaire on misinformation feedback during ChatGPT, based on the research by Arechar et al. (2023) and Pennycook et al. (2021). The sample item is "ChatGPT may provide contradictory answers". |
| | positive affect toward ChatGPT | Adapted from the questionnaire on positive affection during ChatGPT based on the research of Kern et al. (2015). The sample item is "I feel very amused when using ChatGPT". |
| | expectation belief | Adapted from the questionnaire on expectancy belief based on the research of Ye et al. (2022). A sample item is "I expect ChatGPT to help me answer questions that I don't understand." |
| | Learning avoidance motivation | Adapted from the questionnaire on avoidance motivation based on the research of Ye et al. (2023). A sample item is "When I'm doing homework, I tend to opt for the simplest methods available." |
| | accuracy motivation | It was designed based on the study of Arechar et al. (2023) and included one question in this study (e.g., "How important is the accuracy of the answers when I decide whether to accept and adopt the answers provided by ChatGPT?"). |
| | self-leadership | It was designed based on the study of Hauschildt et al. (2012), and included 10 items in this study (e.g., "I hold myself accountable for achieving the established goal.") |
| Background information | sex | What is your sex? (0 = female, 1 = male) |
| | age | What is your age? |
| | grade | What is your grade? |
| | ChatGPT use experience | What is your use experience with ChatGPT? 1=Less than 1 month; 2=1 to 3 months; 3=3 to 6 months ;4=More than 6 months |

Appendix B. The core characteristics of each algorithm

Seven machine learning techniques are employed to assess and compare model performance. The core features of each algorithm are as follows:

**Logistic Regression** is a common machine learning algorithm for binary classification, mapping features to probability space for predictions. It excels on linearly separable datasets and demonstrates computational efficiency (Dreiseitl & Ohno-Machado, 2002). However, its assumption of a linear relationship between features and outcomes may limit its performance in more complex scenarios.

**Naïve Bayes** is a simple form of Bayesian models, belonging to supervised probabilistic machine learning algorithms. It applies Bayes' theorem with the 'naïve' assumption of conditional independence among features, given the class variable's value (Pedregosa et al., 2011). This approach allows for easy implementation without extensive hyperparameter tuning. Consequently, Naïve Bayes models are effective for large-scale datasets (Merghadi et al., 2020).

**Decision tree** is a commonly used method for classification and regression tasks (Mengash, 2020) due to its simplicity and interpretability. It represents a tree-shaped structure where each internal node signifies a feature, each branch denotes a decision based on that feature, and each leaf node signifies the outcome or decision. However, a significant drawback of this algorithm is its susceptibility to overfitting (Tomasevic et al., 2020).

**Random Forest** is a supervised classification algorithm and is classified as an ensemble method. It utilizes Decision Tree models, where each tree is trained on a subset of the data independently sampled using bootstrapping (Breiman et al., 2003). The term "forest" in Random Forest signifies the collection of many individual Decision Tree models. Through voting or averaging predictions, it produces robust results and effectively mitigates overfitting (Merghadi et al., 2020).

**K-Nearest Neighbors** (KNN) operates by analyzing the k-closest training instances in the feature space. In classification tasks, KNN outputs class membership probabilities, indicating the uncertainty in assigning items to classes. Essentially, classification is based on the majority vote of the k nearest neighbors, where k is typically a small positive integer. When k equals 1, the object is assigned to the class of its single nearest neighbor (Chen et al., 2020). While KNN is straightforward and adaptable to different data types, it may underperform with high-dimensional or noisy data and is computationally expensive for large datasets (Merghadi et al., 2020).

**Light Gradient Boosting Machine** (LGBM) is a robust ensemble learning technique rooted in gradient boosting trees. It amalgamates the predictions of multiple weak learners to generate a formidable learner (Ke et al., 2017). By leveraging histograms, it expedites decision tree training, employs a leaf-wise growth strategy to enhance model complexity, and enhances generalization capability through histogram bias correction. These strategies collectively enable efficient and precise model training and prediction.

**Artificial Neural Network** (ANN) is designed to address complex problems by simulating the human brain (Al-Alawi et al., 2023; Khanna et al., 2016). They comprise interconnected units that input, process, and output data for further analysis. The output of each unit is determined by the weighted sum of its inputs, with these weights indicating signal strength (Tomasevic et al., 2020). However, neural networks are often considered black box models (i.e., intricate architectures and opaque decision-making mechanisms), making the internal decision-making process difficult to interpret (Dobson, 2023).

Appendix C. Machine learning model performances

Table S1 presents accuracy scores for both the training and testing sets, along with the performance metrics of different machine learning algorithms. Notably, Random Forest, LightGBM, and ANN demonstrate superior testing accuracy compared to other algorithms. This indicates their practical efficacy in accurately predicting university students' academic help-seeking preferences based on all available features.

Regarding the results of different machine learning algorithms, Random Forest exhibited superior performance with an AUC of 0.747, Accuracy of 0.656, Precision of 0.647, and F1 score of 0.647 among the algorithms tested. However, Logistic Regression, Naive Bayes, Decision Tree, and KNN had lower performance metrics across AUC, Accuracy, Precision, Recall, and F1 score, indicating that none of these algorithms adequately predict the features of university students' help-seeking preferences.

For LightGBM and ANN, both algorithms yielded similar results, all above 0.6. However, the Receiver Operating Characteristic (ROC) curves (see Fig. S1) revealed that the AUC score for LightGBM (0.724) surpassed that of ANN (0.711). Additionally, considering that LightGBM performs better when handling structured data and datasets with fewer features, and it also provides stronger model interpretability compared to neural networks, which are often considered black box models with harder-to-explain internal structures. We consider applying LightGBM for analysis in this study.

Overall, we found that Random Forest and LightGBM exhibit notable algorithmic advantages in evaluating the importance of each factor in predicting academic help-seeking preference. This reflects that the Random Forest is relatively straightforward to fine-tune and exhibits robustness to parameter adjustments when compared to other algorithms (Lv et al., 2022). LightGBM, being a highly efficient gradient boosting decision tree (GBDT), strategically partitions continuous eigenvalues into k intervals with division points chosen from among the k values. Consequently, it excels in terms of both training speed and space efficiency compared to other algorithms (Ju et al., 2019). Hence, there may be a reason why these two algorithms won the comparisons.

It is noteworthy that, to enhance the accuracy and performance of the model, we integrated the feature importance from both LightGBM and Random Forest algorithms. We have developed a combined model to predict university students' preferences for using Google or ChatGPT when seeking help online. Importantly, given a notable quantitative disparity in feature importance between the two algorithms, we normalized the maximum values for both LightGBM and Random Forest to 100. Subsequently, the average value of feature importance across both algorithms was calculated to predict the key predictive factors when college students seek academic help online.

Table S1. Accuracy scores of the training set and testing set, and machine learning algorithm

performances

| Techniques | Train | Test | AUC | Accuracy | Precision | Recall | F1 score |
|---|---|---|---|---|---|---|---|
| Logistic Regression | 0.649 | 0.604 | 0.666 | 0.604 | 0.590 | 0.612 | 0.601 |
| Naive Bayes | 0.608 | 0.547 | 0.598 | 0.547 | 0.528 | 0.676 | 0.593 |
| Decision Tree | 1.000 | 0.607 | 0.608 | 0.607 | 0.589 | 0.640 | 0.614 |
| Random Forest | **1.000** | **0.656** | **0.747** | **0.656** | **0.647** | **0.647** | **0.647** |
| KNN | 0.754 | 0.628 | 0.703 | 0.628 | 0.630 | 0.576 | 0.602 |
| LightGBM | **0.787** | **0.635** | **0.724** | **0.635** | **0.634** | **0.627** | **0.615** |
| ANN | 0.988 | 0.625 | 0.711 | 0.625 | 0.610 | 0.640 | 0.625 |

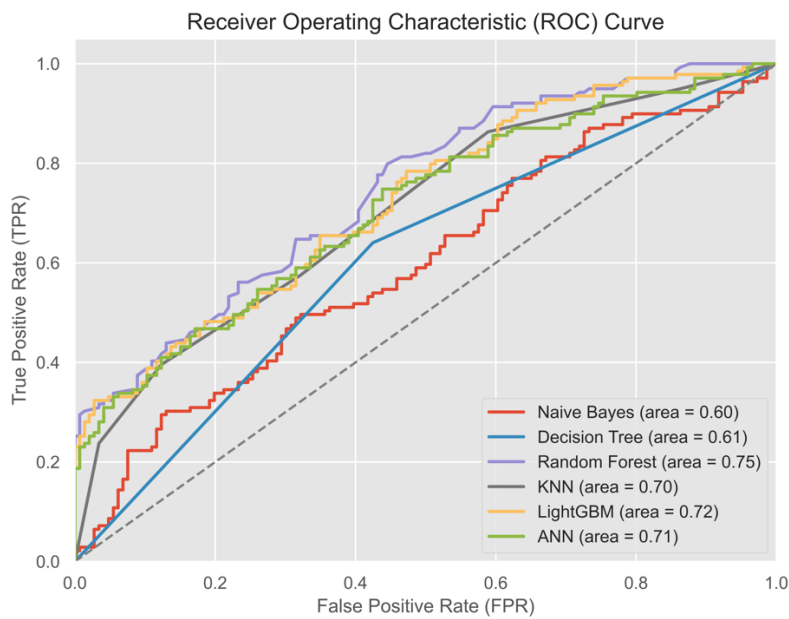

Fig. S1. ROC curves for machine learning model